\documentclass[aps,prl,reprint,superscriptaddress,nofootinbib,floatfix]{revtex4-2}

\usepackage{amsmath}
\usepackage{amssymb}
\usepackage{bm}
\usepackage{mathtools}
\usepackage{graphicx}
\usepackage{xcolor}
\usepackage{hyperref}
\hypersetup{
	colorlinks=true,
	linkcolor=blue,
	citecolor=blue,
	urlcolor=blue
}

\providecommand{\doi}[1]{\href{https://doi.org/#1}{\nolinkurl{doi:#1}}}
\newcommand{\arxiv}[1]{\href{https://arxiv.org/abs/#1}{\nolinkurl{arXiv:#1}}}

\newcommand{\Cov}{\operatorname{Cov}}
\newcommand{\Var}{\operatorname{Var}}

\begin{document}

\title{Optimal all-angle reconstruction of the Hellings--Downs curve}

\author{Jing-Hong Han}
\affiliation{Department of Applied Physics, College of Science, China Agricultural University, Qinghua East Road, Beijing 100083, People's Republic of China}

\author{Zhi-Chao Zhao}
\email{zhaozc@cau.edu.cn}
\thanks{Corresponding author}
\affiliation{Department of Applied Physics, College of Science, China Agricultural University, Qinghua East Road, Beijing 100083, People's Republic of China}

\begin{abstract}
Pulsar timing arrays (PTAs) detect nanohertz gravitational waves through spatial correlations between the timing residuals of different pulsars. For an isotropic, unpolarized stochastic background in general relativity, the ensemble-mean correlation follows the Hellings--Downs (HD) curve; measuring this angular pattern tests the gravitational-wave origin of the signal. Standard bin-by-bin reconstructions optimize the weights within each angular bin separately. We reconstruct the curve jointly using the full pulsar-pair covariance, retaining a free amplitude in every bin. The resulting all-angle best linear unbiased estimator minimizes the variance of every bin value and every linear combination of bins.
Applied to the public NANOGrav 15 yr data products, our method reduces the bin standard deviations by up to 12.9\%, with a median reduction of 9.5\%. For a future Square Kilometre Array Observatory (SKAO)-like PTA, the predicted reduction reaches 36.4\%, with a median of 29.6\%, enabling significantly more precise measurements of the gravitational-wave background.
\end{abstract}

\maketitle

\paragraph{Introduction.---}
Pulsar timing arrays (PTAs) search for nanohertz gravitational waves through correlations induced in the timing residuals of different pulsars~\cite{Sazhin1978,Detweiler1979,FosterBacker1990}. For a stationary, isotropic, and unpolarized stochastic tensor background in general relativity, the ensemble-mean correlation depends only on the angular separation of a pulsar pair and is proportional to the Hellings--Downs (HD) curve~\cite{HellingsDowns1983}. Detection of this angular correlation is crucial for establishing the gravitational-wave origin of a common signal~\cite{AllenRomanoFAQ2024,AllenRomano2025PRL}. NANOGrav, EPTA/InPTA, PPTA, and CPTA have reported spatial correlations consistent with the HD curve~\cite{Agazie2023GWB,Antoniadis2023EPTA,Reardon2023PPTA,Xu2023CPTA}. Beyond establishing the presence of a gravitational-wave background, the angular correlations can test alternative gravitational-wave polarizations and departures from statistical isotropy~\cite{ChamberlinSiemens2012,Gair2014,Gair2015,Agazie2023Anisotropy}. The constraining power of these tests is related to how precisely the angular correlation can be reconstructed.

The HD curve is commonly reconstructed by grouping pulsar pairs according to angular separation and estimating one curve value in each bin~\cite{Agazie2023GWB,Johnson2024Methods,AllenRomano2023PRD,AllenRomano2025PRL}. The uncertainty of this reconstruction contains pulsar variance, associated with using a finite set of pulsars at specific sky locations, and cosmic variance, arising because the array observes a single realization of the stochastic gravitational-wave background~\cite{AllenVariance2023,AllenRomano2023PRD}. Existing bin-by-bin minimum-variance estimators account for correlations among the pulsar-pair measurements and optimally combine their frequency-domain information, but determine the weights for each bin using only the pairs assigned to that bin and their within-bin covariance~\cite{AllenRomano2023PRD,AllenRomano2025PRL}. Cross-bin correlations therefore offer a further source of precision for reconstructing the angular pattern of the background.

In this Letter, we exploit this unused cross-bin information without changing the binned parameterization. Treating every bin amplitude as an independent parameter, we determine the weights of all bin estimators jointly from every pulsar pair and the full covariance. The resulting all-angle estimator has minimum covariance among all jointly unbiased linear estimators. For the same data and binning, the covariance matrices \(\bm\Sigma_{\rm all}\) and \(\bm\Sigma_{\rm bin}\) of the all-angle and bin-by-bin reconstructions, respectively, satisfy
\[
\bm\Sigma_{\rm all}\preceq\bm\Sigma_{\rm bin},
\]
where \(\preceq\) denotes the positive-semidefinite order. Consequently, no individual bin estimate or linear combination of the reconstructed curve can have a larger variance. This guarantee is achieved without smoothing across angular bins, reducing the number of bin parameters, or imposing relations among their amplitudes. For finite arrays, the gain exploits geometry-dependent correlations; in the uniformly sampled, zero-bin-width limit, both reconstructions attain the same cosmic-variance floor~\cite{AllenVariance2023,AllenRomano2023PRD,AllenRomano2025PRL}, as shown in the End Matter. Applied to the public NANOGrav 15 yr data products, the method reduces the standard deviations in all 15 bins, with a median reduction of \(9.5\%\). A SKAO-like forecast gives a median reduction of \(29.6\%\), showing that the reconstruction method becomes more important as PTA measurements improve.

\paragraph{All-angle reconstruction of the HD curve.---}

An HD reconstruction typically uses correlations formed from a finite set of pulsars at specific sky locations, with the pulsar pairs grouped into finite-width angular bins~\cite{AllenRomano2023PRD,AllenRomano2025PRL}. Correlations formed from different pulsar pairs have a non-diagonal covariance, even when the pairs belong to different angular bins~\cite{AllenVariance2023,AllenRomano2023PRD,AllenRomano2025PRL}. This covariance does not disappear as the number of pulsars increases. For infinitely many uniformly distributed pulsars and vanishing bin width, the covariance between estimates at angular separations \(\gamma\) and \(\gamma'\) approaches the cosmic covariance \(s(\gamma,\gamma')\)~\cite{AllenRomano2023PRD}. By contrast, for a finite set of pulsars at fixed sky locations, the covariance retains array-specific structure beyond the cosmic-covariance limit. We use this structure to construct an all-angle reconstruction from measurements across angular separations, while keeping the mean correlation in each bin as an independent parameter.

For \(N_\gamma\) angular bins, let \(\bm\mu=(\mu_1,\ldots,\mu_{N_\gamma})^t\) collect the free parameters specifying the mean angular correlation at the representative bin angles. These amplitudes are inferred from a data vector \(\bm Z\) built from pulsar-pair correlations. For current frequency-combined products, e.g. NANOGrav data products used below, \(\bm Z\) contains one correlation per pulsar pair~\cite{Agazie2023GWB,Johnson2024Methods}. For future high-precision PTA analyses, the frequency information can instead be retained explicitly, with each pulsar pair contributing multiple Fourier-domain pair products to \(\bm Z\), indexed by one frequency for each pulsar~\cite{AllenRomano2025PRL}. The ensemble mean and covariance of \(\bm Z\) take the form
\begin{equation}
	\langle\bm Z\rangle=\bm R\bm\mu,
	\qquad
	\Cov(\bm Z)=\bm C.
	\label{eq:joint_response}
\end{equation}
Here \(\bm R\) contains the frequency response and the standard HD variation across each angular bin~\cite{HellingsDowns1983,AllenRomano2025PRL}. The explicit frequency-domain response is given in the End Matter.

To reconstruct the curve without imposing relations among the bins, a linear estimator must recover every possible vector \(\bm\mu\) in expectation. Writing the reconstructed bin values as \(\widehat{\bm\mu}=\bm W\bm Z\), where \(\bm W\) contains the weights applied to the pulsar-pair correlations, this joint unbiasedness requires
\begin{equation}
	\bm W\bm R=\bm I_{N_\gamma}.
	\label{eq:unbiased_condition}
\end{equation}
Equation~\eqref{eq:unbiased_condition} ensures an unbiased reconstruction, \(\langle\widehat{\bm\mu}\rangle=\bm\mu\), with each bin amplitude treated as an independent parameter.
Let \(i\) label the entries of \(\bm Z\), and let \(B_s\) be the set of indices whose pulsar pairs lie in angular bin \(s\). The bin-by-bin reconstruction restricts the estimate of \(\mu_s\) to the entries in \(B_s\)~\cite{AllenRomano2023PRD,AllenRomano2025PRL}. This means
\begin{equation}
	W_{s;i}=0
	\qquad \text{for }i\notin B_s.
	\label{eq:bin_support}
\end{equation}
However, this restriction is unnecessary because Eq.~\eqref{eq:unbiased_condition} already guarantees unbiased reconstruction while retaining each bin amplitude as an independent parameter. Removing it would allow pulsar-pair measurements across all angular separations to help cancel correlated fluctuations.

We define the all-angle reconstruction by removing this restriction. We seek a single weight matrix \(\bm W\) that, subject to Eq.~\eqref{eq:unbiased_condition}, simultaneously minimizes the variance of every reconstructed bin value. Since \(\widehat{\bm\mu}=\bm W\bm Z\) and \(\Cov(\bm Z)=\bm C\), the covariance of the full reconstruction is \(\bm W\bm C\bm W^\dagger\), whose \(s\)th diagonal element is \(\operatorname{Var}(\widehat{\mu}_s)\). This joint minimum-variance problem yields the generalized least-squares weights~\cite{Tegmark1997,vanderVaart1998}
\begin{equation}
	\bm W_{\rm all}
	=\bigl(\bm R^\dagger\bm C^{-1}\bm R\bigr)^{-1}
	\bm R^\dagger\bm C^{-1}.
	\label{eq:all_angle_weights}
\end{equation}
Applying these weights to the data gives the all-angle estimate and its covariance
\begin{equation}
	\widehat{\bm\mu}_{\rm all}=\bm W_{\rm all}\bm Z,
	\qquad
	\bm\Sigma_{\rm all}
	\equiv\Cov(\widehat{\bm\mu}_{\rm all})
	=\bigl(\bm R^\dagger\bm C^{-1}\bm R\bigr)^{-1}.
	\label{eq:all_angle_estimator_covariance}
\end{equation}

The optimality of these weights can now be shown directly. Let
\(\widehat{\bm\mu}'=\bm W'\bm Z\) be any other jointly unbiased
linear reconstruction, with covariance
\(\bm\Sigma'=\bm W'\bm C\bm W'^\dagger\). Since both \(\bm W'\) and
\(\bm W_{\rm all}\)
satisfy Eq.~\eqref{eq:unbiased_condition}, their difference
\(\Delta\bm W\equiv\bm W'-\bm W_{\rm all}\) obeys
\(\Delta\bm W\bm R=0\). Equation~\eqref{eq:all_angle_weights} then
gives
\(\bm W_{\rm all}\bm C\Delta\bm W^\dagger
=\bm\Sigma_{\rm all}(\Delta\bm W\bm R)^\dagger=0\).
Its Hermitian conjugate also vanishes, so
\begin{equation}
	\bm\Sigma'
	=
	\bm\Sigma_{\rm all}
	+
	\Delta\bm W\bm C\Delta\bm W^\dagger
	\succeq
	\bm\Sigma_{\rm all}.
	\label{eq:BLUE_order}
\end{equation}
Because \(\bm C\) is a covariance matrix, it is positive semidefinite,
and so is \(\Delta\bm W\bm C\Delta\bm W^\dagger\). The \(s\)th
diagonal element of Eq.~\eqref{eq:BLUE_order} therefore gives
\[
	\Var(\widehat\mu'_s)
	-
	\Var(\widehat\mu_{{\rm all},s})
	=
	\bigl(\Delta\bm W\bm C\Delta\bm W^\dagger\bigr)_{ss}
	\geq 0
\]
for every bin \(s\). Thus no other jointly unbiased linear
reconstruction can have a smaller variance in any bin.
Equation~\eqref{eq:BLUE_order} therefore establishes the optimality of
\(\bm W_{\rm all}\) among all jointly unbiased linear reconstructions.

The bin-by-bin reconstruction, with weights \(\bm W_{\rm bin}\), is one such reconstruction,
with the additional restriction in Eq.~\eqref{eq:bin_support}. Setting
\(\bm W'=\bm W_{\rm bin}\) in Eq.~\eqref{eq:BLUE_order} gives
\begin{equation}
	\bm\Sigma_{\rm all}\preceq\bm\Sigma_{\rm bin}.
	\label{eq:Sigma_order}
\end{equation}
The \(s\)th diagonal element of this relation gives
\(\Var(\widehat\mu_{{\rm all},s})
\leq\Var(\widehat\mu_{{\rm bin},s})\) for every bin \(s\).

A central use of the reconstructed bins is to distinguish the HD curve
from competing angular patterns, including monopolar and dipolar
correlations~
\cite{AllenRomano2023PRD,Agazie2023GWB,Johnson2024Methods,Sardesai2023}
and correlations produced by alternative gravitational-wave
polarizations or modified propagation~
\cite{ChamberlinSiemens2012,Gair2015,BernardoNg2023}.
Such a test combines the bins as
\(\bm a^t\widehat{\bm\mu}\), where the \(N_\gamma\)-component vector
\(\bm a\) weights their predicted differences against their correlated
uncertainties~
\cite{Anholm2009,Chamberlin2015,Vigeland2018,Sardesai2023}.
For any fixed \(\bm a\), Eq.~\eqref{eq:BLUE_order} gives
\[
	\Var(\bm a^t\widehat{\bm\mu}_{\rm all})
	\leq
	\Var(\bm a^t\widehat{\bm\mu}').
\]
The all-angle reconstruction therefore has minimum variance for every
linear probe of the angular correlations among all jointly unbiased
linear reconstructions.

For a specified comparison, let \(\bm\mu_{\rm HD}\) denote the true
binned HD mean and \(\bm\mu_{\rm alt}\) the binned prediction of a
specified competing pattern, and define
\(\delta\bm\mu\equiv\bm\mu_{\rm alt}-\bm\mu_{\rm HD}\).
For a reconstruction with covariance \(\bm\Sigma\), the predicted
contrast between the two binned patterns along \(\bm a\) is
\(\bm a^t\delta\bm\mu\), while the corresponding linear statistic
\(\bm a^t\widehat{\bm\mu}\) has variance
\(\bm a^t\bm\Sigma\bm a\). The largest squared signal-to-noise ratio is
\(\delta\bm\mu^t\bm\Sigma^{-1}\delta\bm\mu\), attained for
\(\bm a\propto\bm\Sigma^{-1}\delta\bm\mu\).
Equation~\eqref{eq:BLUE_order} therefore implies
\begin{equation}
	\delta\bm\mu^t\bm\Sigma_{\rm all}^{-1}\delta\bm\mu
	\geq
	\delta\bm\mu^t\bm\Sigma'^{-1}\delta\bm\mu.
	\label{eq:departure_sensitivity}
\end{equation}
Thus, when the standard HD curve is the true mean, among all jointly
unbiased linear reconstructions, the all-angle reconstruction yields
the largest optimized signal-to-noise ratio for separating it from any
specified competing binned angular pattern.

\begin{figure}[tbp]
	\centering
	\includegraphics[width=\linewidth]
	{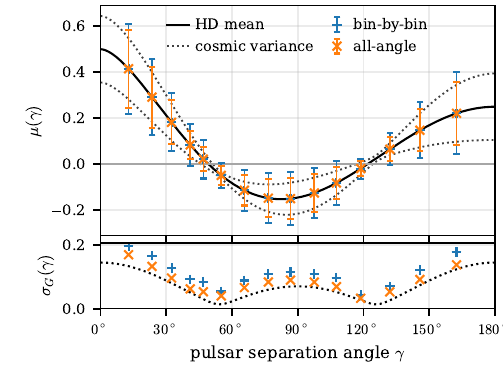}
	\caption{
		Binned HD correlation uncertainties for the NANOGrav pulsar sky locations in the single-frequency, negligible-noise limit. The solid curve is the HD mean \(\mu_{\rm u}(\gamma)\) at pair separation \(\gamma\), normalized by \(\mu_{\rm u}(0)=1/2\); \(\sigma_G\) denotes the standard deviation obtained from the geometric covariance \(\bm G\). Blue ``\(+\)'' and orange ``\(\times\)'' symbols show the bin-by-bin and all-angle \(1\sigma_G\) ranges about this curve. Black dotted curves give the pointwise cosmic-variance limit for infinitely many uniformly distributed pulsars~\cite{RoebberHolder2017,AllenVariance2023,AllenRomano2023PRD,AllenRomano2025PRL}. The lower panel compares the corresponding standard deviations.
	}
	\label{fig:ng15_geometry}
\end{figure}

To isolate the angular-geometric origin of the gain, consider the
signal-dominated, single-frequency limit with negligible pulsar noise~
\cite{AllenVariance2023,AllenRomano2023PRD,AllenRomano2025PRL}.
Measurements from pulsar pairs \(ab\) and \(cd\) are generally
correlated even when the two pairs lie in different angular bins.
Apart from an overall frequency-dependent normalization, their
geometry-dependent covariance is
\begin{equation}
	G_{ab;cd}=\mu_{ac}\mu_{bd}+\mu_{ad}\mu_{bc}.
	\label{eq:geometry_covariance_kernel}
\end{equation}
Here \(\mu_{xy}=\mu_u(\gamma_{xy})(1+\delta_{xy})\), including the pulsar-term contribution, so that \(\mu_{xx}=1\) in our normalization.  
 Evaluating \(\bm G\)
for the NANOGrav pulsar locations and using its normalization from the End Matter in
Eq.~\eqref{eq:all_angle_estimator_covariance} gives the orange all-angle
uncertainties in Fig.~\ref{fig:ng15_geometry}, while the blue
uncertainties show the bin-by-bin result. The bin standard
deviations decrease by up to \(37.3\%\), with a median reduction of
\(24.8\%\), bringing the all-angle reconstruction closer to the dotted
cosmic-variance envelope~
\cite{RoebberHolder2017,AllenVariance2023,AllenRomano2023PRD,AllenRomano2025PRL}.

The construction in
Eqs.~\eqref{eq:joint_response}--\eqref{eq:Sigma_order} is not restricted
to a single frequency. When multiple Fourier frequencies are retained,
each pulsar pair contributes frequency-resolved entries to \(\bm Z\),
while \(\bm R\) and \(\bm C\) contain their full angular-frequency
response and covariance. The same joint optimization then determines
the weights across both pulsar pairs and Fourier frequencies~
\cite{AllenRomano2025PRL}; the corresponding matrix-valued effective-frequency description is given in the End Matter.

\paragraph{Current and future PTAs.---}

\begin{figure}[tbp]
	\centering
	\includegraphics[width=\linewidth]
	{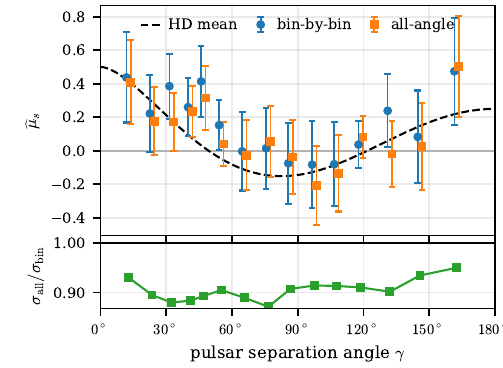}
	\caption{
		Reconstructions of the public NANOGrav 15 yr pair correlations. Blue circles and orange squares show the bin-by-bin and all-angle results, respectively, with \(1\sigma\) error bars. The black dashed line is the standard HD curve, and the lower panel shows \(\sigma_{{\rm all},s}/\sigma_{{\rm bin},s}\).
	}
	\label{fig:ng15_hd_response}
\end{figure}

To demonstrate the gain from cross-bin information in current PTA
data, Fig.~\ref{fig:ng15_hd_response} compares the bin-by-bin and all-angle reconstructions of the NANOGrav 15 yr data. We reproduce the bin-by-bin result reported in Ref.~\cite{Agazie2023GWB} and use the same configuration for the all-angle reconstruction. Blue circles and orange squares show the bin-by-bin and all-angle results, respectively. The all-angle reconstruction reduces the standard deviations in all 15 bins by
as much as \(12.9\%\), with a median reduction of \(9.5\%\).

Both reconstructions are unbiased for the same independent bin
amplitudes, as independently confirmed by the simulations described
below. For a particular realization, however, the distinct weights
combine the correlated fluctuations differently and generally yield
different central values. The offsets in
Fig.~\ref{fig:ng15_hd_response} therefore reflect the observed
realization rather than estimator bias, while the smaller error bars
arise solely from the enlarged weight space.

\begin{figure}[tbp]
	\centering
	\includegraphics[width=\linewidth]
	{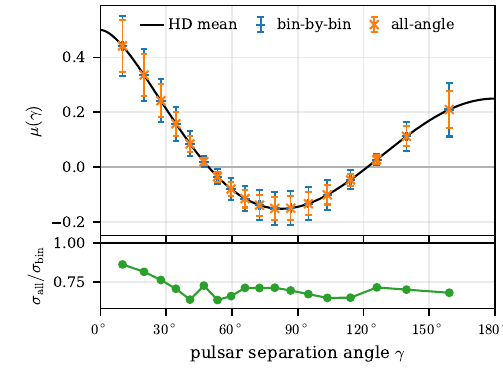}
	\caption{
		Broadband SKAO-like simulations. Blue ``\(+\)'' and orange ``\(\times\)'' symbols show the sample means of the bin-by-bin and all-angle reconstructions from \(10^5\) Gaussian realizations, with \(1\sigma\) sample standard deviations. The lower panel shows \(\sigma_{{\rm all},s}/\sigma_{{\rm bin},s}\).
	}
	\label{fig:skao_broadband}
\end{figure}

The all-angle reconstruction can offer substantially larger gains for
future high-precision PTAs. As an example, we consider a broadband
SKAO-like forecast using the 174-pulsar array proposed in
Ref.~\cite{Shannon2025SKAOPTA} and the signal and noise model described
in the End Matter.
The forecast retains 16 positive Fourier frequencies, with the all-angle weights optimized jointly across pulsar pairs and frequencies.
For this configuration, \(10^5\) Gaussian
realizations are analyzed with both reconstructions. Figure~
\ref{fig:skao_broadband} shows the resulting sample means and standard
deviations. The all-angle standard deviations decrease by up to
\(36.4\%\), with a median reduction of \(29.6\%\). The median reduction
is more than three times that obtained for the current NANOGrav
products, illustrating the growing value of cross-bin information as
PTA precision improves.

The sample means of both reconstructions recover the input HD curve.
Their sample covariances agree within Monte Carlo accuracy with the
analytic predictions
\(\bm\Sigma_{\rm bin}=\bm W_{\rm bin}\bm C\bm W_{\rm bin}^\dagger\) and
\(\bm\Sigma_{\rm all}=(\bm R^\dagger\bm C^{-1}\bm R)^{-1}\) from
Eq.~\eqref{eq:all_angle_estimator_covariance}, confirming both
unbiasedness and the predicted variance reduction.

\paragraph{Discussion.---}

In this Letter, we have developed an all-angle reconstruction that
jointly infers the binned HD curve from all pulsar-pair measurements and
their full covariance. Compared with the bin-by-bin
reconstruction, it reduces the bin standard deviations by up to
\(12.9\%\) for the public NANOGrav 15 yr data products and by up to
\(36.4\%\) for a future SKAO-like PTA. Simulations of \(10^5\)
realizations recover the input HD curve and reproduce the analytic
covariance, numerically validating the unbiasedness and predicted
precision of the reconstruction. 

These gains arise because a finite set
of pulsars at specific sky locations does not possess the full rotational
symmetry of a continuous uniform-sky average~
\cite{AllenVariance2023,AllenRomano2023PRD}. The covariance between
measurements at different separations can therefore be used to construct
combinations that leave every bin mean unchanged while canceling part of
the fluctuations in a target-bin estimate. In the idealized limit of
infinitely many uniformly distributed pulsars, vanishing bin width, and negligible pulsar noise, rotational symmetry removes this opportunity. Realistic forecasts
remain at \(\mathcal O(10^2)\) well-timed pulsars even for multi-decade
SKAO observations~\cite{Rosado2015,Shannon2025SKAOPTA}, placing this
limit far beyond the planned SKAO era. This demonstrates the continuing importance of all-angle reconstruction for future high-precision PTAs.

All pulsar pairs can also be used to reconstruct the angular correlation
in a Legendre basis, with the harmonic coefficients inferred from the
full array~\cite{AllenRomanoHarmonic2024,Nay2024}. With infinitely many modes,
the Legendre expansion may represent an arbitrary angular
correlation function. However, practical
harmonic analyses retain only a small number of multipoles supported by
the data, providing a compact and smooth representation of the angular
correlation. This truncation reduces the parameter dimension but links
the inferred correlations at different \(\gamma\). In contrast, the
all-angle reconstruction developed here separates all-pair covariance
weighting from angular compression. It directly estimates the binned
observables reported by PTA analyses and retains every bin amplitude as
an independent parameter. Relations among angular bins are therefore
inferred from the data rather than imposed by the parameterization. This
retains sensitivity to deviations from the HD curve, whether they extend
across the angular range or are localized in individual bins.

The all-angle reconstruction extends naturally to non-Gaussian or
discrete-source backgrounds, because its optimal weights are determined
by the mean response and covariance of the pair measurements~
\cite{AllenValtolina2024,Becsy2022}. When signal and noise parameters
are inferred from the data, their uncertainty can be incorporated by
evaluating the reconstruction over posterior samples~\cite{Vigeland2018,Johnson2024Methods}. Future PTA
analyses that retain the full pair-correlation covariance and its
frequency dependence will allow all available cross-bin information to
be exploited. As PTA precision improves, these gains will sharpen
measurements of the spatial correlation pattern that establishes the
gravitational-wave origin of a common signal and probes the properties
of the nanohertz background.

\begin{acknowledgments}
This work is supported by the National Key Research and Development
Program of China Grant No.~2021YFC2203001, the 2115 Talent Development
Program of China Agricultural University, and the High-performance
Computing Platform of China Agricultural University.
\end{acknowledgments}

\bibliographystyle{apsrev4-2}
\bibliography{refs}

\begin{thebibliography}{37}%
\makeatletter
\providecommand \@ifxundefined [1]{%
 \@ifx{#1\undefined}
}%
\providecommand \@ifnum [1]{%
 \ifnum #1\expandafter \@firstoftwo
 \else \expandafter \@secondoftwo
 \fi
}%
\providecommand \@ifx [1]{%
 \ifx #1\expandafter \@firstoftwo
 \else \expandafter \@secondoftwo
 \fi
}%
\providecommand \natexlab [1]{#1}%
\providecommand \enquote  [1]{``#1''}%
\providecommand \bibnamefont  [1]{#1}%
\providecommand \bibfnamefont [1]{#1}%
\providecommand \citenamefont [1]{#1}%
\providecommand \href@noop [0]{\@secondoftwo}%
\providecommand \href [0]{\begingroup \@sanitize@url \@href}%
\providecommand \@href[1]{\@@startlink{#1}\@@href}%
\providecommand \@@href[1]{\endgroup#1\@@endlink}%
\providecommand \@sanitize@url [0]{\catcode `\\12\catcode `\$12\catcode
  `\&12\catcode `\#12\catcode `\^12\catcode `\_12\catcode `\%12\relax}%
\providecommand \@@startlink[1]{}%
\providecommand \@@endlink[0]{}%
\providecommand \url  [0]{\begingroup\@sanitize@url \@url }%
\providecommand \@url [1]{\endgroup\@href {#1}{\urlprefix }}%
\providecommand \urlprefix  [0]{URL }%
\providecommand \Eprint [0]{\href }%
\providecommand \doibase [0]{https://doi.org/}%
\providecommand \selectlanguage [0]{\@gobble}%
\providecommand \bibinfo  [0]{\@secondoftwo}%
\providecommand \bibfield  [0]{\@secondoftwo}%
\providecommand \translation [1]{[#1]}%
\providecommand \BibitemOpen [0]{}%
\providecommand \bibitemStop [0]{}%
\providecommand \bibitemNoStop [0]{.\EOS\space}%
\providecommand \EOS [0]{\spacefactor3000\relax}%
\providecommand \BibitemShut  [1]{\csname bibitem#1\endcsname}%
\let\auto@bib@innerbib\@empty
\bibitem [{\citenamefont {Sazhin}(1978)}]{Sazhin1978}%
  \BibitemOpen
  \bibfield  {author} {\bibinfo {author} {\bibfnamefont {M.~V.}\ \bibnamefont
  {Sazhin}},\ }\href@noop {} {\bibfield  {journal} {\bibinfo  {journal} {Sov.
  Astron.}\ }\textbf {\bibinfo {volume} {22}},\ \bibinfo {pages} {36} (\bibinfo
  {year} {1978})}\BibitemShut {NoStop}%
\bibitem [{\citenamefont {Detweiler}(1979)}]{Detweiler1979}%
  \BibitemOpen
  \bibfield  {author} {\bibinfo {author} {\bibfnamefont {S.~L.}\ \bibnamefont
  {Detweiler}},\ }\href {https://doi.org/10.1086/157593} {\bibfield  {journal}
  {\bibinfo  {journal} {Astrophys. J.}\ }\textbf {\bibinfo {volume} {234}},\
  \bibinfo {pages} {1100} (\bibinfo {year} {1979})}\BibitemShut {NoStop}%
\bibitem [{\citenamefont {Foster}\ and\ \citenamefont
  {Backer}(1990)}]{FosterBacker1990}%
  \BibitemOpen
  \bibfield  {author} {\bibinfo {author} {\bibfnamefont {R.~S.}\ \bibnamefont
  {Foster}}\ and\ \bibinfo {author} {\bibfnamefont {D.~C.}\ \bibnamefont
  {Backer}},\ }\href {https://doi.org/10.1086/169195} {\bibfield  {journal}
  {\bibinfo  {journal} {Astrophys. J.}\ }\textbf {\bibinfo {volume} {361}},\
  \bibinfo {pages} {300} (\bibinfo {year} {1990})}\BibitemShut {NoStop}%
\bibitem [{\citenamefont {Hellings}\ and\ \citenamefont
  {Downs}(1983)}]{HellingsDowns1983}%
  \BibitemOpen
  \bibfield  {author} {\bibinfo {author} {\bibfnamefont {R.~W.}\ \bibnamefont
  {Hellings}}\ and\ \bibinfo {author} {\bibfnamefont {G.~S.}\ \bibnamefont
  {Downs}},\ }\href {https://doi.org/10.1086/183954} {\bibfield  {journal}
  {\bibinfo  {journal} {Astrophys. J. Lett.}\ }\textbf {\bibinfo {volume}
  {265}},\ \bibinfo {pages} {L39} (\bibinfo {year} {1983})}\BibitemShut
  {NoStop}%
\bibitem [{\citenamefont {Romano}\ and\ \citenamefont
  {Allen}(2024)}]{AllenRomanoFAQ2024}%
  \BibitemOpen
  \bibfield  {author} {\bibinfo {author} {\bibfnamefont {J.~D.}\ \bibnamefont
  {Romano}}\ and\ \bibinfo {author} {\bibfnamefont {B.}~\bibnamefont {Allen}},\
  }\href {https://doi.org/10.1088/1361-6382/ad4c4c} {\bibfield  {journal}
  {\bibinfo  {journal} {Class. Quant. Grav.}\ }\textbf {\bibinfo {volume}
  {41}},\ \bibinfo {pages} {175008} (\bibinfo {year} {2024})},\ \Eprint
  {https://arxiv.org/abs/2308.05847} {arXiv:2308.05847 [gr-qc]} \BibitemShut
  {NoStop}%
\bibitem [{\citenamefont {Allen}\ and\ \citenamefont
  {Romano}(2025)}]{AllenRomano2025PRL}%
  \BibitemOpen
  \bibfield  {author} {\bibinfo {author} {\bibfnamefont {B.}~\bibnamefont
  {Allen}}\ and\ \bibinfo {author} {\bibfnamefont {J.~D.}\ \bibnamefont
  {Romano}},\ }\href {https://doi.org/10.1103/PhysRevLett.134.031401}
  {\bibfield  {journal} {\bibinfo  {journal} {Phys. Rev. Lett.}\ }\textbf
  {\bibinfo {volume} {134}},\ \bibinfo {pages} {031401} (\bibinfo {year}
  {2025})},\ \Eprint {https://arxiv.org/abs/2407.10968} {arXiv:2407.10968
  [gr-qc]} \BibitemShut {NoStop}%
\bibitem [{\citenamefont {Agazie}\ \emph
  {et~al.}(2023{\natexlab{a}})\citenamefont {Agazie} \emph
  {et~al.}}]{Agazie2023GWB}%
  \BibitemOpen
  \bibfield  {author} {\bibinfo {author} {\bibfnamefont {G.}~\bibnamefont
  {Agazie}} \emph {et~al.} (\bibinfo {collaboration} {NANOGrav}),\ }\href
  {https://doi.org/10.3847/2041-8213/acdac6} {\bibfield  {journal} {\bibinfo
  {journal} {Astrophys. J. Lett.}\ }\textbf {\bibinfo {volume} {951}},\
  \bibinfo {pages} {L8} (\bibinfo {year} {2023}{\natexlab{a}})},\ \Eprint
  {https://arxiv.org/abs/2306.16213} {arXiv:2306.16213 [astro-ph.HE]}
  \BibitemShut {NoStop}%
\bibitem [{\citenamefont {Antoniadis}\ \emph {et~al.}(2023)\citenamefont
  {Antoniadis} \emph {et~al.}}]{Antoniadis2023EPTA}%
  \BibitemOpen
  \bibfield  {author} {\bibinfo {author} {\bibfnamefont {J.}~\bibnamefont
  {Antoniadis}} \emph {et~al.} (\bibinfo {collaboration} {EPTA, InPTA}),\
  }\href {https://doi.org/10.1051/0004-6361/202346844} {\bibfield  {journal}
  {\bibinfo  {journal} {Astron. Astrophys.}\ }\textbf {\bibinfo {volume}
  {678}},\ \bibinfo {pages} {A50} (\bibinfo {year} {2023})},\ \Eprint
  {https://arxiv.org/abs/2306.16214} {arXiv:2306.16214 [astro-ph.HE]}
  \BibitemShut {NoStop}%
\bibitem [{\citenamefont {Reardon}\ \emph {et~al.}(2023)\citenamefont {Reardon}
  \emph {et~al.}}]{Reardon2023PPTA}%
  \BibitemOpen
  \bibfield  {author} {\bibinfo {author} {\bibfnamefont {D.~J.}\ \bibnamefont
  {Reardon}} \emph {et~al.},\ }\href {https://doi.org/10.3847/2041-8213/acdd02}
  {\bibfield  {journal} {\bibinfo  {journal} {Astrophys. J. Lett.}\ }\textbf
  {\bibinfo {volume} {951}},\ \bibinfo {pages} {L6} (\bibinfo {year} {2023})},\
  \Eprint {https://arxiv.org/abs/2306.16215} {arXiv:2306.16215 [astro-ph.HE]}
  \BibitemShut {NoStop}%
\bibitem [{\citenamefont {Xu}\ \emph {et~al.}(2023)\citenamefont {Xu} \emph
  {et~al.}}]{Xu2023CPTA}%
  \BibitemOpen
  \bibfield  {author} {\bibinfo {author} {\bibfnamefont {H.}~\bibnamefont {Xu}}
  \emph {et~al.},\ }\href {https://doi.org/10.1088/1674-4527/acdfa5} {\bibfield
   {journal} {\bibinfo  {journal} {Res. Astron. Astrophys.}\ }\textbf {\bibinfo
  {volume} {23}},\ \bibinfo {pages} {075024} (\bibinfo {year} {2023})},\
  \Eprint {https://arxiv.org/abs/2306.16216} {arXiv:2306.16216 [astro-ph.HE]}
  \BibitemShut {NoStop}%
\bibitem [{\citenamefont {Chamberlin}\ and\ \citenamefont
  {Siemens}(2012)}]{ChamberlinSiemens2012}%
  \BibitemOpen
  \bibfield  {author} {\bibinfo {author} {\bibfnamefont {S.~J.}\ \bibnamefont
  {Chamberlin}}\ and\ \bibinfo {author} {\bibfnamefont {X.}~\bibnamefont
  {Siemens}},\ }\href {https://doi.org/10.1103/PhysRevD.85.082001} {\bibfield
  {journal} {\bibinfo  {journal} {Phys. Rev. D}\ }\textbf {\bibinfo {volume}
  {85}},\ \bibinfo {pages} {082001} (\bibinfo {year} {2012})},\ \Eprint
  {https://arxiv.org/abs/1111.5661} {arXiv:1111.5661 [astro-ph.HE]}
  \BibitemShut {NoStop}%
\bibitem [{\citenamefont {Gair}\ \emph {et~al.}(2014)\citenamefont {Gair},
  \citenamefont {Romano}, \citenamefont {Taylor},\ and\ \citenamefont
  {Mingarelli}}]{Gair2014}%
  \BibitemOpen
  \bibfield  {author} {\bibinfo {author} {\bibfnamefont {J.}~\bibnamefont
  {Gair}}, \bibinfo {author} {\bibfnamefont {J.~D.}\ \bibnamefont {Romano}},
  \bibinfo {author} {\bibfnamefont {S.}~\bibnamefont {Taylor}},\ and\ \bibinfo
  {author} {\bibfnamefont {C.~M.~F.}\ \bibnamefont {Mingarelli}},\ }\href
  {https://doi.org/10.1103/PhysRevD.90.082001} {\bibfield  {journal} {\bibinfo
  {journal} {Phys. Rev. D}\ }\textbf {\bibinfo {volume} {90}},\ \bibinfo
  {pages} {082001} (\bibinfo {year} {2014})},\ \Eprint
  {https://arxiv.org/abs/1406.4664} {arXiv:1406.4664 [gr-qc]} \BibitemShut
  {NoStop}%
\bibitem [{\citenamefont {Gair}\ \emph {et~al.}(2015)\citenamefont {Gair},
  \citenamefont {Romano},\ and\ \citenamefont {Taylor}}]{Gair2015}%
  \BibitemOpen
  \bibfield  {author} {\bibinfo {author} {\bibfnamefont {J.~R.}\ \bibnamefont
  {Gair}}, \bibinfo {author} {\bibfnamefont {J.~D.}\ \bibnamefont {Romano}},\
  and\ \bibinfo {author} {\bibfnamefont {S.~R.}\ \bibnamefont {Taylor}},\
  }\href {https://doi.org/10.1103/PhysRevD.92.102003} {\bibfield  {journal}
  {\bibinfo  {journal} {Phys. Rev. D}\ }\textbf {\bibinfo {volume} {92}},\
  \bibinfo {pages} {102003} (\bibinfo {year} {2015})},\ \Eprint
  {https://arxiv.org/abs/1506.08668} {arXiv:1506.08668 [gr-qc]} \BibitemShut
  {NoStop}%
\bibitem [{\citenamefont {Agazie}\ \emph
  {et~al.}(2023{\natexlab{b}})\citenamefont {Agazie} \emph
  {et~al.}}]{Agazie2023Anisotropy}%
  \BibitemOpen
  \bibfield  {author} {\bibinfo {author} {\bibfnamefont {G.}~\bibnamefont
  {Agazie}} \emph {et~al.} (\bibinfo {collaboration} {NANOGrav}),\ }\href
  {https://doi.org/10.3847/2041-8213/acf4fd} {\bibfield  {journal} {\bibinfo
  {journal} {Astrophys. J. Lett.}\ }\textbf {\bibinfo {volume} {956}},\
  \bibinfo {pages} {L3} (\bibinfo {year} {2023}{\natexlab{b}})},\ \Eprint
  {https://arxiv.org/abs/2306.16221} {arXiv:2306.16221 [astro-ph.HE]}
  \BibitemShut {NoStop}%
\bibitem [{\citenamefont {Johnson}\ \emph {et~al.}(2024)\citenamefont {Johnson}
  \emph {et~al.}}]{Johnson2024Methods}%
  \BibitemOpen
  \bibfield  {author} {\bibinfo {author} {\bibfnamefont {A.~D.}\ \bibnamefont
  {Johnson}} \emph {et~al.} (\bibinfo {collaboration} {NANOGrav}),\ }\href
  {https://doi.org/10.1103/PhysRevD.109.103012} {\bibfield  {journal} {\bibinfo
   {journal} {Phys. Rev. D}\ }\textbf {\bibinfo {volume} {109}},\ \bibinfo
  {pages} {103012} (\bibinfo {year} {2024})},\ \Eprint
  {https://arxiv.org/abs/2306.16223} {arXiv:2306.16223 [astro-ph.HE]}
  \BibitemShut {NoStop}%
\bibitem [{\citenamefont {Allen}\ and\ \citenamefont
  {Romano}(2023)}]{AllenRomano2023PRD}%
  \BibitemOpen
  \bibfield  {author} {\bibinfo {author} {\bibfnamefont {B.}~\bibnamefont
  {Allen}}\ and\ \bibinfo {author} {\bibfnamefont {J.~D.}\ \bibnamefont
  {Romano}},\ }\href {https://doi.org/10.1103/PhysRevD.108.043026} {\bibfield
  {journal} {\bibinfo  {journal} {Phys. Rev. D}\ }\textbf {\bibinfo {volume}
  {108}},\ \bibinfo {pages} {043026} (\bibinfo {year} {2023})},\ \Eprint
  {https://arxiv.org/abs/2208.07230} {arXiv:2208.07230 [gr-qc]} \BibitemShut
  {NoStop}%
\bibitem [{\citenamefont {Allen}(2023)}]{AllenVariance2023}%
  \BibitemOpen
  \bibfield  {author} {\bibinfo {author} {\bibfnamefont {B.}~\bibnamefont
  {Allen}},\ }\href {https://doi.org/10.1103/PhysRevD.107.043018} {\bibfield
  {journal} {\bibinfo  {journal} {Phys. Rev. D}\ }\textbf {\bibinfo {volume}
  {107}},\ \bibinfo {pages} {043018} (\bibinfo {year} {2023})},\ \Eprint
  {https://arxiv.org/abs/2205.05637} {arXiv:2205.05637 [gr-qc]} \BibitemShut
  {NoStop}%
\bibitem [{\citenamefont {Tegmark}(1997)}]{Tegmark1997}%
  \BibitemOpen
  \bibfield  {author} {\bibinfo {author} {\bibfnamefont {M.}~\bibnamefont
  {Tegmark}},\ }\href {https://doi.org/10.1103/PhysRevD.55.5895} {\bibfield
  {journal} {\bibinfo  {journal} {Phys. Rev. D}\ }\textbf {\bibinfo {volume}
  {55}},\ \bibinfo {pages} {5895} (\bibinfo {year} {1997})},\ \Eprint
  {https://arxiv.org/abs/astro-ph/9611174} {arXiv:astro-ph/9611174}
  \BibitemShut {NoStop}%
\bibitem [{\citenamefont {van~der Vaart}(1998)}]{vanderVaart1998}%
  \BibitemOpen
  \bibfield  {author} {\bibinfo {author} {\bibfnamefont {A.~W.}\ \bibnamefont
  {van~der Vaart}},\ }\href {https://doi.org/10.1017/CBO9780511802256} {\emph
  {\bibinfo {title} {{Asymptotic Statistics}}}},\ Cambridge Series in
  Statistical and Probabilistic Mathematics\ (\bibinfo  {publisher} {Cambridge
  University Press},\ \bibinfo {year} {1998})\BibitemShut {NoStop}%
\bibitem [{\citenamefont {Sardesai}\ \emph {et~al.}(2023)\citenamefont
  {Sardesai}, \citenamefont {Vigeland}, \citenamefont {Gersbach},\ and\
  \citenamefont {Taylor}}]{Sardesai2023}%
  \BibitemOpen
  \bibfield  {author} {\bibinfo {author} {\bibfnamefont {S.~C.}\ \bibnamefont
  {Sardesai}}, \bibinfo {author} {\bibfnamefont {S.~J.}\ \bibnamefont
  {Vigeland}}, \bibinfo {author} {\bibfnamefont {K.~A.}\ \bibnamefont
  {Gersbach}},\ and\ \bibinfo {author} {\bibfnamefont {S.~R.}\ \bibnamefont
  {Taylor}},\ }\href {https://doi.org/10.1103/PhysRevD.108.124081} {\bibfield
  {journal} {\bibinfo  {journal} {Phys. Rev. D}\ }\textbf {\bibinfo {volume}
  {108}},\ \bibinfo {pages} {124081} (\bibinfo {year} {2023})},\ \Eprint
  {https://arxiv.org/abs/2303.09615} {arXiv:2303.09615 [astro-ph.IM]}
  \BibitemShut {NoStop}%
\bibitem [{\citenamefont {Bernardo}\ and\ \citenamefont
  {Ng}(2023)}]{BernardoNg2023}%
  \BibitemOpen
  \bibfield  {author} {\bibinfo {author} {\bibfnamefont {R.~C.}\ \bibnamefont
  {Bernardo}}\ and\ \bibinfo {author} {\bibfnamefont {K.-W.}\ \bibnamefont
  {Ng}},\ }\href {https://doi.org/10.1103/PhysRevD.107.L101502} {\bibfield
  {journal} {\bibinfo  {journal} {Phys. Rev. D}\ }\textbf {\bibinfo {volume}
  {107}},\ \bibinfo {pages} {L101502} (\bibinfo {year} {2023})},\ \Eprint
  {https://arxiv.org/abs/2302.11796} {arXiv:2302.11796 [gr-qc]} \BibitemShut
  {NoStop}%
\bibitem [{\citenamefont {Anholm}\ \emph {et~al.}(2009)\citenamefont {Anholm},
  \citenamefont {Ballmer}, \citenamefont {Creighton}, \citenamefont {Price},\
  and\ \citenamefont {Siemens}}]{Anholm2009}%
  \BibitemOpen
  \bibfield  {author} {\bibinfo {author} {\bibfnamefont {M.}~\bibnamefont
  {Anholm}}, \bibinfo {author} {\bibfnamefont {S.}~\bibnamefont {Ballmer}},
  \bibinfo {author} {\bibfnamefont {J.~D.~E.}\ \bibnamefont {Creighton}},
  \bibinfo {author} {\bibfnamefont {L.~R.}\ \bibnamefont {Price}},\ and\
  \bibinfo {author} {\bibfnamefont {X.}~\bibnamefont {Siemens}},\ }\href
  {https://doi.org/10.1103/PhysRevD.79.084030} {\bibfield  {journal} {\bibinfo
  {journal} {Phys. Rev. D}\ }\textbf {\bibinfo {volume} {79}},\ \bibinfo
  {pages} {084030} (\bibinfo {year} {2009})},\ \Eprint
  {https://arxiv.org/abs/0809.0701} {arXiv:0809.0701 [gr-qc]} \BibitemShut
  {NoStop}%
\bibitem [{\citenamefont {Chamberlin}\ \emph {et~al.}(2015)\citenamefont
  {Chamberlin}, \citenamefont {Creighton}, \citenamefont {Demorest},
  \citenamefont {Ellis}, \citenamefont {Price}, \citenamefont {Romano},\ and\
  \citenamefont {Siemens}}]{Chamberlin2015}%
  \BibitemOpen
  \bibfield  {author} {\bibinfo {author} {\bibfnamefont {S.~J.}\ \bibnamefont
  {Chamberlin}}, \bibinfo {author} {\bibfnamefont {J.~D.~E.}\ \bibnamefont
  {Creighton}}, \bibinfo {author} {\bibfnamefont {P.~B.}\ \bibnamefont
  {Demorest}}, \bibinfo {author} {\bibfnamefont {J.~A.}\ \bibnamefont {Ellis}},
  \bibinfo {author} {\bibfnamefont {L.~R.}\ \bibnamefont {Price}}, \bibinfo
  {author} {\bibfnamefont {J.~D.}\ \bibnamefont {Romano}},\ and\ \bibinfo
  {author} {\bibfnamefont {X.}~\bibnamefont {Siemens}},\ }\href
  {https://doi.org/10.1103/PhysRevD.91.044048} {\bibfield  {journal} {\bibinfo
  {journal} {Phys. Rev. D}\ }\textbf {\bibinfo {volume} {91}},\ \bibinfo
  {pages} {044048} (\bibinfo {year} {2015})},\ \Eprint
  {https://arxiv.org/abs/1410.8256} {arXiv:1410.8256 [astro-ph.IM]}
  \BibitemShut {NoStop}%
\bibitem [{\citenamefont {Vigeland}\ \emph {et~al.}(2018)\citenamefont
  {Vigeland}, \citenamefont {Islo}, \citenamefont {Taylor},\ and\ \citenamefont
  {Ellis}}]{Vigeland2018}%
  \BibitemOpen
  \bibfield  {author} {\bibinfo {author} {\bibfnamefont {S.~J.}\ \bibnamefont
  {Vigeland}}, \bibinfo {author} {\bibfnamefont {K.}~\bibnamefont {Islo}},
  \bibinfo {author} {\bibfnamefont {S.~R.}\ \bibnamefont {Taylor}},\ and\
  \bibinfo {author} {\bibfnamefont {J.~A.}\ \bibnamefont {Ellis}},\ }\href
  {https://doi.org/10.1103/PhysRevD.98.044003} {\bibfield  {journal} {\bibinfo
  {journal} {Phys. Rev. D}\ }\textbf {\bibinfo {volume} {98}},\ \bibinfo
  {pages} {044003} (\bibinfo {year} {2018})},\ \Eprint
  {https://arxiv.org/abs/1805.12188} {arXiv:1805.12188 [astro-ph.IM]}
  \BibitemShut {NoStop}%
\bibitem [{\citenamefont {Roebber}\ and\ \citenamefont
  {Holder}(2017)}]{RoebberHolder2017}%
  \BibitemOpen
  \bibfield  {author} {\bibinfo {author} {\bibfnamefont {E.}~\bibnamefont
  {Roebber}}\ and\ \bibinfo {author} {\bibfnamefont {G.}~\bibnamefont
  {Holder}},\ }\href {https://doi.org/10.3847/1538-4357/835/1/21} {\bibfield
  {journal} {\bibinfo  {journal} {Astrophys. J.}\ }\textbf {\bibinfo {volume}
  {835}},\ \bibinfo {pages} {21} (\bibinfo {year} {2017})},\ \Eprint
  {https://arxiv.org/abs/1609.06758} {arXiv:1609.06758 [astro-ph.CO]}
  \BibitemShut {NoStop}%
\bibitem [{\citenamefont {Shannon}\ \emph {et~al.}(2025)\citenamefont {Shannon}
  \emph {et~al.}}]{Shannon2025SKAOPTA}%
  \BibitemOpen
  \bibfield  {author} {\bibinfo {author} {\bibfnamefont {R.~M.}\ \bibnamefont
  {Shannon}} \emph {et~al.} (\bibinfo {collaboration} {SKAO Pulsar Science
  Working Group}),\ }\href {https://doi.org/10.33232/001c.154243} {\bibfield
  {journal} {\bibinfo  {journal} {Open J. Astrophys.}\ }\textbf {\bibinfo
  {volume} {8}},\ \bibinfo {pages} {54243} (\bibinfo {year} {2025})},\ \Eprint
  {https://arxiv.org/abs/2512.16163} {arXiv:2512.16163 [astro-ph.HE]}
  \BibitemShut {NoStop}%
\bibitem [{\citenamefont {Rosado}\ \emph {et~al.}(2015)\citenamefont {Rosado},
  \citenamefont {Sesana},\ and\ \citenamefont {Gair}}]{Rosado2015}%
  \BibitemOpen
  \bibfield  {author} {\bibinfo {author} {\bibfnamefont {P.~A.}\ \bibnamefont
  {Rosado}}, \bibinfo {author} {\bibfnamefont {A.}~\bibnamefont {Sesana}},\
  and\ \bibinfo {author} {\bibfnamefont {J.}~\bibnamefont {Gair}},\ }\href
  {https://doi.org/10.1093/mnras/stv1098} {\bibfield  {journal} {\bibinfo
  {journal} {Mon. Not. Roy. Astron. Soc.}\ }\textbf {\bibinfo {volume} {451}},\
  \bibinfo {pages} {2417} (\bibinfo {year} {2015})},\ \Eprint
  {https://arxiv.org/abs/1503.04803} {arXiv:1503.04803 [astro-ph.HE]}
  \BibitemShut {NoStop}%
\bibitem [{\citenamefont {Allen}\ and\ \citenamefont
  {Romano}(2024)}]{AllenRomanoHarmonic2024}%
  \BibitemOpen
  \bibfield  {author} {\bibinfo {author} {\bibfnamefont {B.}~\bibnamefont
  {Allen}}\ and\ \bibinfo {author} {\bibfnamefont {J.~D.}\ \bibnamefont
  {Romano}},\ }\href@noop {} {\bibinfo {title} {{Harmonic spectrum of pulsar
  timing array angular correlations}}} (\bibinfo {year} {2024}),\ \Eprint
  {https://arxiv.org/abs/2412.14852} {arXiv:2412.14852 [gr-qc]} \BibitemShut
  {NoStop}%
\bibitem [{\citenamefont {Nay}\ \emph {et~al.}(2024)\citenamefont {Nay},
  \citenamefont {Boddy}, \citenamefont {Smith},\ and\ \citenamefont
  {Mingarelli}}]{Nay2024}%
  \BibitemOpen
  \bibfield  {author} {\bibinfo {author} {\bibfnamefont {J.}~\bibnamefont
  {Nay}}, \bibinfo {author} {\bibfnamefont {K.~K.}\ \bibnamefont {Boddy}},
  \bibinfo {author} {\bibfnamefont {T.~L.}\ \bibnamefont {Smith}},\ and\
  \bibinfo {author} {\bibfnamefont {C.~M.~F.}\ \bibnamefont {Mingarelli}},\
  }\href {https://doi.org/10.1103/PhysRevD.110.044062} {\bibfield  {journal}
  {\bibinfo  {journal} {Phys. Rev. D}\ }\textbf {\bibinfo {volume} {110}},\
  \bibinfo {pages} {044062} (\bibinfo {year} {2024})},\ \Eprint
  {https://arxiv.org/abs/2306.06168} {arXiv:2306.06168 [gr-qc]} \BibitemShut
  {NoStop}%
\bibitem [{\citenamefont {Allen}\ and\ \citenamefont
  {Valtolina}(2024)}]{AllenValtolina2024}%
  \BibitemOpen
  \bibfield  {author} {\bibinfo {author} {\bibfnamefont {B.}~\bibnamefont
  {Allen}}\ and\ \bibinfo {author} {\bibfnamefont {S.}~\bibnamefont
  {Valtolina}},\ }\href {https://doi.org/10.1103/PhysRevD.109.083038}
  {\bibfield  {journal} {\bibinfo  {journal} {Phys. Rev. D}\ }\textbf {\bibinfo
  {volume} {109}},\ \bibinfo {pages} {083038} (\bibinfo {year} {2024})},\
  \Eprint {https://arxiv.org/abs/2401.14329} {arXiv:2401.14329 [gr-qc]}
  \BibitemShut {NoStop}%
\bibitem [{\citenamefont {B{\'e}csy}\ \emph {et~al.}(2022)\citenamefont
  {B{\'e}csy}, \citenamefont {Cornish},\ and\ \citenamefont
  {Kelley}}]{Becsy2022}%
  \BibitemOpen
  \bibfield  {author} {\bibinfo {author} {\bibfnamefont {B.}~\bibnamefont
  {B{\'e}csy}}, \bibinfo {author} {\bibfnamefont {N.~J.}\ \bibnamefont
  {Cornish}},\ and\ \bibinfo {author} {\bibfnamefont {L.~Z.}\ \bibnamefont
  {Kelley}},\ }\href {https://doi.org/10.3847/1538-4357/aca1b2} {\bibfield
  {journal} {\bibinfo  {journal} {Astrophys. J.}\ }\textbf {\bibinfo {volume}
  {941}},\ \bibinfo {pages} {119} (\bibinfo {year} {2022})},\ \Eprint
  {https://arxiv.org/abs/2207.01607} {arXiv:2207.01607 [astro-ph.HE]}
  \BibitemShut {NoStop}%
\bibitem [{\citenamefont {Isserlis}(1918)}]{Isserlis1918}%
  \BibitemOpen
  \bibfield  {author} {\bibinfo {author} {\bibfnamefont {L.}~\bibnamefont
  {Isserlis}},\ }\href {https://doi.org/10.1093/biomet/12.1-2.134} {\bibfield
  {journal} {\bibinfo  {journal} {Biometrika}\ }\textbf {\bibinfo {volume}
  {12}},\ \bibinfo {pages} {134} (\bibinfo {year} {1918})}\BibitemShut
  {NoStop}%
\bibitem [{\citenamefont {Moore}(1920)}]{Moore1920}%
  \BibitemOpen
  \bibfield  {author} {\bibinfo {author} {\bibfnamefont {E.~H.}\ \bibnamefont
  {Moore}},\ }\href {https://doi.org/10.1090/S0002-9904-1920-03322-7}
  {\bibfield  {journal} {\bibinfo  {journal} {Bull. Am. Math. Soc.}\ }\textbf
  {\bibinfo {volume} {26}},\ \bibinfo {pages} {394} (\bibinfo {year}
  {1920})}\BibitemShut {NoStop}%
\bibitem [{\citenamefont {Penrose}(1955)}]{Penrose1955}%
  \BibitemOpen
  \bibfield  {author} {\bibinfo {author} {\bibfnamefont {R.}~\bibnamefont
  {Penrose}},\ }\href {https://doi.org/10.1017/S0305004100030401} {\bibfield
  {journal} {\bibinfo  {journal} {Proc. Cambridge Philos. Soc.}\ }\textbf
  {\bibinfo {volume} {51}},\ \bibinfo {pages} {406} (\bibinfo {year}
  {1955})}\BibitemShut {NoStop}%
\bibitem [{\citenamefont {Spiewak}\ \emph
  {et~al.}(2022{\natexlab{a}})\citenamefont {Spiewak} \emph
  {et~al.}}]{Spiewak2022MeerTime}%
  \BibitemOpen
  \bibfield  {author} {\bibinfo {author} {\bibfnamefont {R.}~\bibnamefont
  {Spiewak}} \emph {et~al.},\ }\href {https://doi.org/10.1017/pasa.2022.19}
  {\bibfield  {journal} {\bibinfo  {journal} {Publ. Astron. Soc. Aust.}\
  }\textbf {\bibinfo {volume} {39}},\ \bibinfo {pages} {e027} (\bibinfo {year}
  {2022}{\natexlab{a}})},\ \Eprint {https://arxiv.org/abs/2204.04115}
  {arXiv:2204.04115 [astro-ph.HE]} \BibitemShut {NoStop}%
\bibitem [{\citenamefont {Spiewak}\ \emph
  {et~al.}(2022{\natexlab{b}})\citenamefont {Spiewak} \emph
  {et~al.}}]{Spiewak2022MeerTimeData}%
  \BibitemOpen
  \bibfield  {author} {\bibinfo {author} {\bibfnamefont {R.}~\bibnamefont
  {Spiewak}} \emph {et~al.},\ }\href {https://doi.org/10.5281/zenodo.5347875}
  {\bibinfo {title} {{MeerTime MSP Census}}} (\bibinfo {year}
  {2022}{\natexlab{b}})\BibitemShut {NoStop}%
\bibitem [{\citenamefont {Allen}(2024)}]{AllenHarmonic2024}%
  \BibitemOpen
  \bibfield  {author} {\bibinfo {author} {\bibfnamefont {B.}~\bibnamefont
  {Allen}},\ }\href {https://doi.org/10.1103/PhysRevD.110.043043} {\bibfield
  {journal} {\bibinfo  {journal} {Phys. Rev. D}\ }\textbf {\bibinfo {volume}
  {110}},\ \bibinfo {pages} {043043} (\bibinfo {year} {2024})},\ \Eprint
  {https://arxiv.org/abs/2404.05677} {arXiv:2404.05677 [gr-qc]} \BibitemShut
  {NoStop}%
\end{thebibliography}%

\appendix

\section*{End Matter}

\subsection{Statistical model and implementation details}
\label{app:statistical_model}

The following derivation adopts the two-sided discrete Fourier convention and Gaussian statistical model of Ref.~\cite{AllenRomano2025PRL}. Let \(Z_a^j\) be the redshift Fourier coefficient of pulsar \(a\), with \(j,k\) labelling the discrete frequencies, and let \(H^{jk}\) and \(P_a^{jk}\) denote the frequency covariance matrices of the stochastic background and the pulsar noise. For a cross-correlated pair \(a<b\), define the symmetrized product \(Z_{ab}^{jk}=(Z_a^jZ_b^k+Z_a^kZ_b^j)/2\). Its mean is
\begin{equation}
	\langle Z_{ab}^{jk}\rangle
	=H^{j,-k}\mu_{\rm u}(\gamma_{ab}),
\end{equation}
where \(\gamma_{ab}\) is the angular separation of pulsars \(a,b\), and \(\mu_{\rm u}\) is the standard HD curve in the normalization \(\mu_{\rm u}(0)=1/2\). Accordingly, \(H^{jk}\) is \(2/3\) times the background frequency matrix in Ref.~\cite{AllenRomano2025PRL}, which uses \(\mu_{\rm u}(0)=1/3\). The covariance between pair products follows from Isserlis' theorem~\cite{Isserlis1918}; the frequency-symmetrized covariance \(C_{ab;cd}^{jk;\ell m}\) is taken from Eqs.~(19) and (20) of Ref.~\cite{AllenRomano2025PRL}.

Let \(\gamma_s\) be the representative angle of bin \(s\), chosen such that \( \mu_u(\gamma_s)\neq0 \). The response appearing in Eq.~\eqref{eq:joint_response} is
\begin{equation}
	R_{ab\,jk,s}
	=
	\begin{cases}
		\displaystyle
		H^{j,-k}
		\frac{\mu_{\rm u}(\gamma_{ab})}{\mu_{\rm u}(\gamma_s)},
		&(ab,j,k)\in B_s,\\[2mm]
		0,&(ab,j,k)\notin B_s.
	\end{cases}
	\label{eq:frequency_response_matrix}
\end{equation}
Thus different bin amplitudes remain independent parameters, while the known variation of the HD response across the finite width of each bin is retained. If \(\bm C\) has redundant null modes with zero mean response, we project both \(\bm Z\) and \(\bm R\) onto its nonzero-eigenvalue subspace. On this subspace, the inverse can equivalently be implemented with the Moore--Penrose pseudoinverse~\cite{Moore1920,Penrose1955}.

For the NANOGrav application, \(\gamma_s\) is the arithmetic mean of the pair separations in bin \(s\). The published internal bin boundaries are retained, while the two outer boundaries are extended to \(0^\circ\) and \(180^\circ\). The public correlation vector and covariance are divided by \(A_{\rm CURN}^2\) and \(A_{\rm CURN}^4\), respectively, where \(A_{\rm CURN}\) is the maximum-likelihood amplitude of the common uncorrelated red-noise process with timing-residual spectral index \(13/3\) in the plotting products~\cite{Johnson2024Methods}. This rescaling changes neither the comparison between estimators nor the ratios shown in Fig.~\ref{fig:ng15_hd_response}. For these frequency-combined data, the nonzero response entries reduce to the angular ratios \(\mu_{\rm u}(\gamma_{ab})/\mu_{\rm u}(\gamma_s)\) in Eq.~\eqref{eq:frequency_response_matrix}.

The pulsar sky positions are taken from the proposed 174-pulsar SKAO
PTA of Ref.~\cite{Shannon2025SKAOPTA}, whose source selection was
derived from the MeerTime millisecond-pulsar census~
\cite{Spiewak2022MeerTime,Spiewak2022MeerTimeData}.
The array is observed for 20 yr at a 14 d cadence with
\(1\,\mu{\rm s}\) white timing-noise standard deviation per pulsar per epoch~
\cite{Shannon2025SKAOPTA}. A Gaussian background with characteristic strain \(h_c(f)=A_{\rm GWB}(f/f_{\rm yr})^\alpha\), where \(f_{\rm yr}=1\,\mathrm{yr}^{-1}\),
\(A_{\rm GWB}=2.4\times10^{-15}\), and \(\alpha=-2/3\), is propagated
through the  pulsar-term and finite-time frequency covariance of
Ref.~\cite{AllenRomano2025PRL}. The calculation retains 16 positive
Fourier frequencies and uses 18 approximately equal-occupancy angular
bins.

\subsection{Effective-frequency description}
\label{app:effective_frequency}

To compare broadband and single-frequency uncertainties, define the geometric response \(\bm R_G\) by \((R_G)_{ab,s}=\mu_{\rm u}(\gamma_{ab})/\mu_{\rm u}(\gamma_s)\) for pairs in bin \(s\), and zero otherwise. The single-frequency geometric benchmark combines two independent, equally normalized real quadratures of one positive Fourier frequency, giving the pair covariance \(\bm G/2\). The binned covariance used in
Fig.~\ref{fig:ng15_geometry} is therefore
\begin{equation}
	\bm\Sigma_G^{\rm all}
	=\bigl(2\bm R_G^t\bm G^{-1}\bm R_G\bigr)^{-1}.
	\label{eq:geometry_fisher_covariance}
\end{equation}
The effective-frequency construction of
Ref.~\cite{AllenRomano2025PRL} becomes matrix valued when the full
binned curve is reconstructed jointly. For broadband data with response \(\bm R_d\) and covariance \(\bm C_d\), let
\begin{equation}
	\bm F_G=(\bm\Sigma_G^{\rm all})^{-1},
	\qquad
	\bm F_d=\bm R_d^\dagger\bm C_d^{-1}\bm R_d,
	\qquad
	\bm\Sigma_d=\bm F_d^{-1}
	\label{eq:data_fisher_covariance}
\end{equation}
denote the single-frequency geometric and broadband all-angle information matrices and the broadband covariance. Define
\begin{equation}
	\bm N_{\rm freq}^{d}
	=\bm F_G^{-1/2}\bm F_d\bm F_G^{-1/2}.
	\label{eq:effective_frequency_operator}
\end{equation}
It follows that
\begin{equation}
	\bm\Sigma_d
	=(\bm\Sigma_G^{\rm all})^{1/2}
	\bigl(\bm N_{\rm freq}^{d}\bigr)^{-1}
	(\bm\Sigma_G^{\rm all})^{1/2}.
	\label{eq:effective_frequency_covariance}
\end{equation}
For one unknown curve value this reduces to the scalar relation \(N_{\rm freq}=\sigma_G^2/\sigma_d^2\) of Ref.~\cite{AllenRomano2025PRL}. For the full curve, different linear combinations can receive different broadband information, so \(\bm N_{\rm freq}^{d}\) need not be proportional to the identity. In the idealized model of Ref.~\cite{AllenRomano2025PRL}, with \(N_{\rm cr}\) independent signal-dominated positive frequencies and orthogonal signal and noise frequency subspaces,
\begin{equation}
	\bm N_{\rm freq}^{d}=N_{\rm cr}\bm I_{N_\gamma},
	\qquad
	\bm\Sigma_d
	=\frac{1}{N_{\rm cr}}\bm\Sigma_G^{\rm all}.
	\label{eq:orthogonal_frequency_main}
\end{equation}
In this idealized model, all-angle weighting changes the angular geometric covariance but not the number of independent signal-dominated frequencies.

\subsection{Cosmic variance for infinitely many uniformly distributed pulsars}
\label{app:cosmic_variance}

Consider a single-frequency, signal-dominated Gaussian background in the limit of negligible measurement noise, infinitely many pulsars, and a uniform sky distribution~\cite{AllenVariance2023,AllenRomano2023PRD,AllenRomano2025PRL}. Let
\(\xi=(\widehat\Omega_1,\widehat\Omega_2)\) denote an ordered pulsar pair with separation
\(\gamma(\xi)=\arccos(\widehat\Omega_1\cdot\widehat\Omega_2)\), and define the fixed-separation shell
\(\mathcal O_\gamma=\{\xi\mid\gamma(\xi)=\gamma\}\).
Let \(Z(\xi)\) be the real, frequency-symmetrized single-frequency pulsar-pair statistic obtained after combining conjugate frequency-pair contributions, with covariance kernel
\(C(\xi,\zeta)=\Cov[Z(\xi),Z(\zeta)]\).
The response and covariance kernel in this limit are real, and all weight functions below are therefore taken to be real. Statistical isotropy implies rotational invariance of the covariance kernel~\cite{AllenVariance2023,AllenHarmonic2024},
\begin{equation}
	C(\mathcal R\xi,\mathcal R\zeta)=C(\xi,\zeta),
	\qquad \mathcal R\in SO(3).
	\label{eq:app_rotation_invariance}
\end{equation}
Here \(\mathcal R\) is an arbitrary three-dimensional rotation.

For uniformly distributed pulsars, let
\[
\mathrm d\nu(\xi)
=
\frac{\mathrm d^2\widehat\Omega_1}{4\pi}
\frac{\mathrm d^2\widehat\Omega_2}{4\pi}
\]
denote the normalized uniform measure on the full space of ordered pulsar pairs.
Grouping the pairs by their separation, this measure can be written as
\[
\mathrm d\nu(\xi)
=
\mathrm d\lambda(\gamma)\,\mathrm d\nu_\gamma(\xi),
\qquad
\mathrm d\lambda(\gamma)
=
\frac12\sin\gamma\,\mathrm d\gamma,
\]
where \(\mathrm d\nu_\gamma\) is the normalized uniform measure on the fixed-separation shell \(\mathcal O_\gamma\). Let \(\mu(\gamma)\) denote the continuous angular-correlation curve in this limit. Since
\(\langle Z(\xi)\rangle=\mu(\gamma(\xi))\), unbiasedness of an estimator of
\(\mu(\gamma_0)\) with weight function \(w\) for every mean curve \(\mu\) requires
\[
\int_{\mathcal O_\gamma}
\mathrm d\nu_\gamma(\xi)\,w(\xi)
=
\delta_{\gamma_0}(\gamma),
\]
where \(\delta_{\gamma_0}\) is the Dirac distribution with respect to
\(\mathrm d\lambda\). The uniform average at the target separation has weight \(w_0(\xi)=\delta_{\gamma_0}(\gamma(\xi))\), which satisfies this condition. Any other weight satisfying the same unbiasedness condition can then be written as
\begin{equation}
	w(\xi)=w_0(\xi)+\eta(\xi),
	\label{eq:app_continuum_weight_decomposition}
\end{equation}
where the additional weight integrates to zero on every separation shell.
\begin{equation}
	\int_{\mathcal O_\gamma}
	\mathrm d\nu_\gamma(\xi)\,\eta(\xi)=0,
	\qquad \forall\gamma.
	\label{eq:app_shell_zero_response}
\end{equation}
The fixed relative HD response within a bin tends to unity as the bin width vanishes, and the condition imposed on finite bins approaches Eq.~\eqref{eq:app_shell_zero_response}. The function \(\eta\) may alter the weights on the target shell and may be nonzero on other shells, but its integral vanishes on each shell. Integrals involving \(w_0\) are understood distributionally; equivalently, one may first work with finite-width bins and then take the zero-width limit. The estimator is written as
\(w\!\cdot\!Z\equiv\int\mathrm d\nu(\xi)\,w(\xi)Z(\xi)\).

For arbitrary weight functions \(u\) and \(v\), define
\[
\langle u,Cv\rangle
\equiv
\int\mathrm d\nu(\xi)\,\mathrm d\nu(\zeta)\,
u(\xi)C(\xi,\zeta)v(\zeta).
\]
Substitution of Eq.~\eqref{eq:app_continuum_weight_decomposition} into the variance gives
\[
\Var(w\cdot Z)
=
\langle w_0,Cw_0\rangle
+2\langle\eta,Cw_0\rangle
+\langle\eta,C\eta\rangle.
\]
Let
\[
\varphi(\zeta)=(Cw_0)(\zeta)
=
\int\mathrm d\nu(\xi)\,C(\zeta,\xi)w_0(\xi).
\]
Equation~\eqref{eq:app_rotation_invariance}, invariance of the measure, and
\(w_0(\mathcal R\xi)=w_0(\xi)\) imply
\begin{align}
	\varphi(\mathcal R\zeta)
	&=
	\int\mathrm d\nu(\xi')\,
	C(\mathcal R\zeta,\mathcal R\xi')w_0(\mathcal R\xi')
	\nonumber\\
	&=
	\int\mathrm d\nu(\xi')\,C(\zeta,\xi')w_0(\xi')
	=\varphi(\zeta).
\end{align}
Any two ordered pulsar pairs with the same separation are related by a common three-dimensional rotation, so \(SO(3)\) acts transitively on each fixed-separation shell. Hence \(\varphi\) is constant on each shell and depends only on the separation,
\(\varphi(\zeta)=\varphi(\gamma(\zeta))\). Together with
Eq.~\eqref{eq:app_shell_zero_response}, this gives
\begin{align}
	\langle\eta,Cw_0\rangle
	&=
	\int\mathrm d\lambda(\gamma)\,\varphi(\gamma)
	\int_{\mathcal O_\gamma}
	\mathrm d\nu_\gamma(\zeta)\,\eta(\zeta)
	\nonumber\\
	&=0.
	\label{eq:app_cross_term_zero}
\end{align}
Therefore
\begin{equation}
	\Var(w\cdot Z)
	=
	\langle w_0,Cw_0\rangle
	+
	\langle\eta,C\eta\rangle
	\geq
	\langle w_0,Cw_0\rangle,
	\label{eq:app_cosmic_variance_bound}
\end{equation}
where the last step follows from positive semidefiniteness of the covariance kernel. Denote by
\(\widetilde{\mu^2}(\gamma)\) the cosmic variance for infinitely many uniformly distributed pulsars and one positive signal-dominated Fourier frequency; we use Eq.~(5) of Ref.~\cite{AllenRomano2025PRL}, rescaled to our normalization \(\mu_{\rm u}(0)=1/2\). In this limit,
\(\langle w_0,Cw_0\rangle=\widetilde{\mu^2}(\gamma_0)\).
Combining \(N_{\rm cr}\) independent signal-dominated frequencies gives~\cite{AllenRomano2025PRL}
\begin{equation}
	\Var_{\rm min}^{\rm all}
	[\widehat\mu(\gamma_0)]
	=
	\frac{\widetilde{\mu^2}(\gamma_0)}{N_{\rm cr}}.
	\label{eq:app_continuum_cosmic_floor}
\end{equation}

If \(C\) is positive definite after removing data null modes, equality in
Eq.~\eqref{eq:app_cosmic_variance_bound} requires \(\eta(\xi)=0\). The optimal weight then becomes the uniform average on the target shell and vanishes pointwise on all non-target shells. If covariance null modes are retained, optimal weights may differ by a null mode, which changes neither the estimator nor its variance. Thus all-angle and fixed-separation reconstructions attain the same cosmic variance in the uniformly sampled, zero-bin-width limit~\cite{AllenVariance2023,AllenRomano2023PRD,AllenRomano2025PRL}.

\end{document}